\documentclass[12pt]{iopart-mod}
\usepackage{epsfig}
\usepackage{amsmath}

\def\cO#1{{\cal O}\left(#1\right)}
\def\as{\alpha_{\textsc{s}}}
\def\ee{e^+e^-}

\def\cf{C_F}
\def\hc{{\cal H}_{\textsc{C}}}
\def\hr{{\cal H}_{\textsc{R}}}
\def\kt{k_t}

\newcommand\hepph[1]{hep-ph/#1}

\newcommand\jhep[3]{{{\it \textsc{JHEP} }{\bf #1} (#2) #3}}

\newcommand\npb[3]{{{\it Nucl. Phys. }{\bf B #1} (#2) #3}}

\newcommand\plb[3]{{{\it Phys. Lett. }{\bf B #1} (#2) #3}}

\newcommand\sjnp[3]{{{\it Sov. J. Nucl. Phys. }{\bf #1} (#2) #3}}
\newcommand\jetp[3]{{{\it Sov. Phys. \textsc{JETP} }{\bf #1} (#2) #3}}

\newcommand\epjdc[3]{{ \it Eur. Phys. J. direct } {\bf {C#1}} (#2) #3}

\begin{document}

\begin{flushright}
  Bicocca--FT--99--31\\
  hep-ph/9910343\\
  October 1999
\end{flushright}

\title{The resummed thrust distribution in DIS}

\author{V. Antonelli\footnote{Talk presented by V. Antonelli at the
UK Phenomenology Workshop on  Collider Physics, September 1999, St.
John's College, Durham.}, M. Dasgupta
  and G. P. Salam}
\address{Dipartimento di Fisica, Universit\`a di Milano-Bicocca \\
  and INFN, Sezione di Milano, Italy}
\begin{abstract}
  We present preliminary results on the resummation of leading and
  next-to-leading logarithms for the thrust distribution in deep
  inelastic scattering. Our predictions, expanded to $\cO{\as^2}$, are
  compared to corresponding results from the Monte Carlo programs
  DISASTER++ and DISENT.
\end{abstract}

\section{Introduction}

For some time now it has been standard practice in $e^+e^-$ reactions
to compare event-shape distributions with resummed perturbative
predictions (see for instance \cite{CTTW,eeExp}). The resummation is
necessary because in the two-jet limit (small values of the shape
variable) the presence of large soft and collinear logarithms spoils
the convergence of the fixed-order calculations.  Such resummed
analyses have led to valuable information about the strong coupling
constant and also about non-perturbative effects.

At HERA, studies of event-shape distributions are being carried out by
both collaborations \cite{HERAdists} but as yet no perturbative
resummed calculations exist for comparison. Here we present
preliminary results on such a calculation \cite{ADS}.

In DIS, event shapes are defined in the current hemisphere of the
Breit frame \cite{BF} to reduce contamination from remnant
fragmentation which is beyond perturbative control. The distribution
of emissions in the current hemisphere ($\hc$) is analogous to that in
a single hemisphere of $e^+e^-$. Differences arise from $\ee$ however
because the momentum of the current quark (the quark struck by the
photon) depends on remnant-hemisphere ($\hr$) emissions through recoil
effects. This necessitates a resummed treatment of the space-like
branching of the incoming parton. Currently such a simultaneous
resummation of the space-like and time-like (double logarithmic)
contributions exists only for jet multiplicities \cite{CDW}, or for
cross sections at large $x$ \cite{CMWlargex}.

For the thrust $T$ (as defined in the following section) the recoil
from remnant hemisphere emissions can be divided into two parts: a
piece from soft and collinear emissions, which gives double logarithms
$\as^n \ln^{2n} (1-T)$, identical to those from half a hemisphere of
$\ee$, and a piece from purely collinear emissions which gives single
logarithms $\as^n \ln^{n} (1-T)$ that can be identified with a
change of scale in the parton distribution. This is outlined in the
next section.

A valuable cross-check of any resummation is to expand it to NLO and
compare the result to that from fixed order Monte Carlo programs such
as DISASTER++ \cite{DR} and DISENT \cite{DT}. This is illustrated in
section~\ref{sec:fixed}.

Finally in section~\ref{sec:future}, we comment on possible future
developments.

\section{The Thrust in DIS}

There are several possible definitions of the thrust in DIS, which
differ according to a choice of axis and the normalisation. We here
consider the thrust with respect to the photon axis, $\vec{n}_\gamma$,
and normalised to $Q/2$,
\begin{equation}     
T = \frac{2}{Q} \sum_{i \in \hc} \vec{k_i}  \cdot \vec{n}_\gamma\,,  
\label{eq:dTQ} 
\end{equation}
where the sum extends over all particles in the current hemisphere,
$H_C$. At lowest order $T=1$, so the region requiring resummation will
be that of $1-T$ close to zero.

Expressing all momenta in terms of their Sudakov components
\begin{equation}\label{eq:suddef}
 k_i\>=\>  \alpha_i P \>+\>\beta_i P' \>+\>k_{t,i}\>,
\qquad\qquad \alpha_i\beta_i\>=\>\vec k_{t,i}^2/Q^2\,,
\end{equation}
where $P=\frac12 Q(1,0,0,-1)$ and $P'=\frac12 Q(1,0,0,1)$ are along
the remnant and current directions respectively, we can write the
thrust as \cite{DWmilan}
\begin{equation} \label{eq:taures}
  1 - T = \sum_{i\in\hr} \beta_i + \sum_{i\in\hc} \alpha_i + \alpha_q\;,
\end{equation}
where the sums now run over emitted particles only and $\alpha_q$ is
the $\alpha$ component of the current quark\footnote{The expression
  must be modified when the current quark goes into the remnant
  hemisphere --- but such a situation is not relevant for small
  $1-T$.}. Dependence on the remnant hemisphere emissions arises both
through the $\sum_{i\in\hr}\beta_i$ term and through
\begin{equation}\label{eq:alphaq}
  \alpha_q = \frac{k^2_{t,q}}{\beta_q}  \simeq \left|- \sum_i {\vec
      k}_{t,i} \right|^2 ,
\end{equation}
(we have made use of the fact that $\beta_q \simeq1$).

We now consider the cross section $\sigma(\tau)$ for $1-T$ to be
smaller than $\tau$. Roughly it will contain virtual corrections from
the exclusion of all emissions whose contribution to \eqref{eq:taures}
is larger than $\tau$.

In the remnant hemisphere we exclude soft and collinear emissions,
$\kt > \beta > \tau$, where the $\kt>\beta$ condition ensures that
particle be in $\hr$.  The integration over $\kt$ and $\beta$ gives a
double logarithm. In $\hc$ we exclude the soft and collinear region
$\kt > \alpha > \tau$.  Again this gives us a double logarithm.  The
exclusion $\alpha_q\simeq k_{t,i}^2 > \tau$ is just a limit on the
maximum emitted transverse momentum: this implies `stopping' the DGLAP
\cite{DGLAP} parton evolution at a scale $\tau Q^2$, which gives single
collinear logarithms.

These aspects are illustrated in the first order result for
$\sigma(\tau)$:
\begin{multline} \label{eq:sigtauas}
  \frac{\sigma(\tau)}{\sigma}
   =  1 - \frac{\as \cf}{2 \pi}
 \left[2 \ln^2 \frac1\tau -
        3 \ln \frac1\tau \right]   \\ -
      \frac{\as}{2\pi}  \frac{\ln 1/\tau}{q(x)}
      \int_x^1  \frac{d z}{z}  
\left[q\left(\frac{x}{z}\right) P_{qq} (z) + 
g\left(\frac{x}{z}\right) P_{qg} (z)  
\right] .
\end{multline}
The first line contains the soft and collinear double logarithms
(which turn out to be identical to the $\ee$ result) from the
conditions on the $\alpha_i$ and $\beta_i$. The second line contains
the collinear logarithm associated with the restriction on the DGLAP
evolution.  Photon-gluon fusion plays a role only through this
collinear logarithm, in the convolution with the gluon distribution.

The actual details of the resummation will be presented in \cite{ADS}.
Schematically, the result is
\begin{equation}
  \label{eq:sigtau}
  \frac{\sigma(\tau)}{\sigma} = 
  \left[1 + \as\, C(x) \right] \frac{q(x,\tau Q^2)}{q(x, Q^2)} \,
  \Sigma(\tau),\qquad \Sigma(\tau) =  e^{ - \frac{\as \cf}{2 \pi}
 \left[2 \ln^2 \frac1\tau -
        3 \ln \frac1\tau \right] + \cO{\as^2\ln^3\frac1\tau}},
\end{equation}
where $\Sigma(\tau)$ is just the corresponding $\ee$ quantity of
\cite{CTTW} and $C(x)$ is a $\tau$-independent coefficient function
which is given in \cite{ADS}.  The ratio $\frac{q(x,\tau Q^2)}{q(x,
  Q^2)}$ comes from the suppression of radiation with $k_t^2 > \tau
Q^2$, as mentioned above.

\section{Comparison with fixed order programs.}
\label{sec:fixed}

\begin{figure}[ht]
   \begin{center}
   \epsfig{file=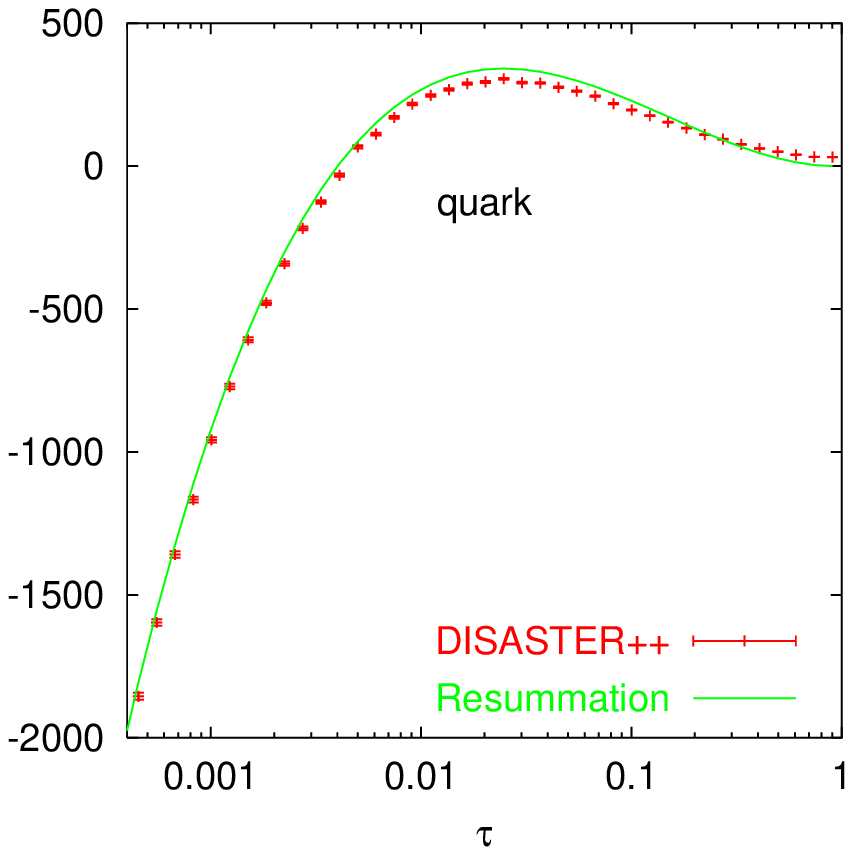,width=0.43\textwidth}\quad
   \epsfig{file=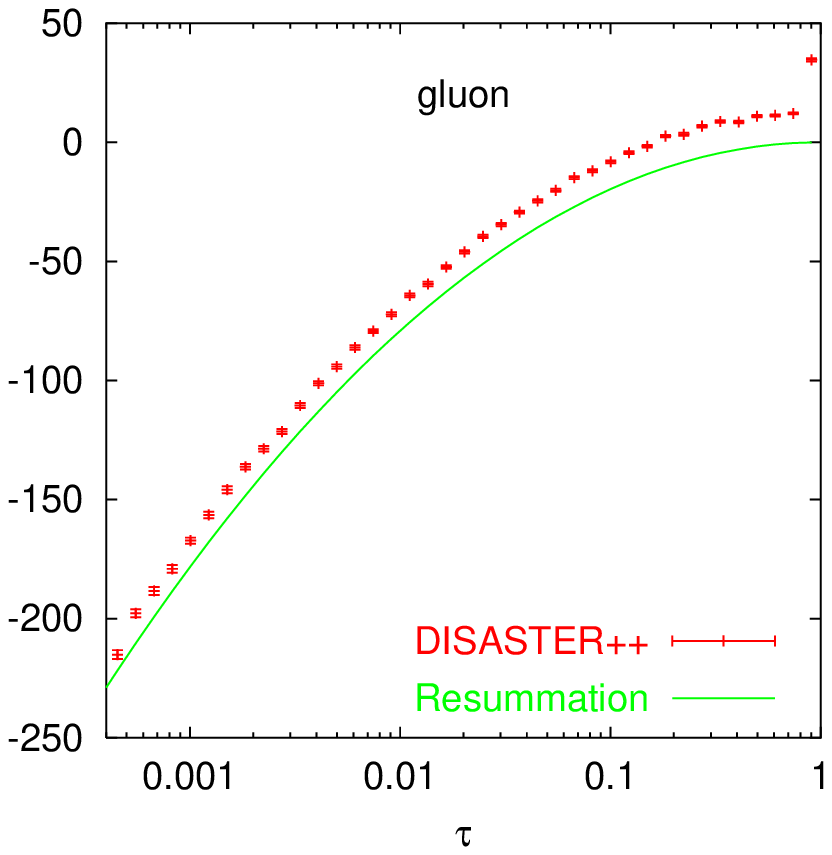,width=0.43\textwidth}\vspace{0.5cm}\\
  \epsfig{file=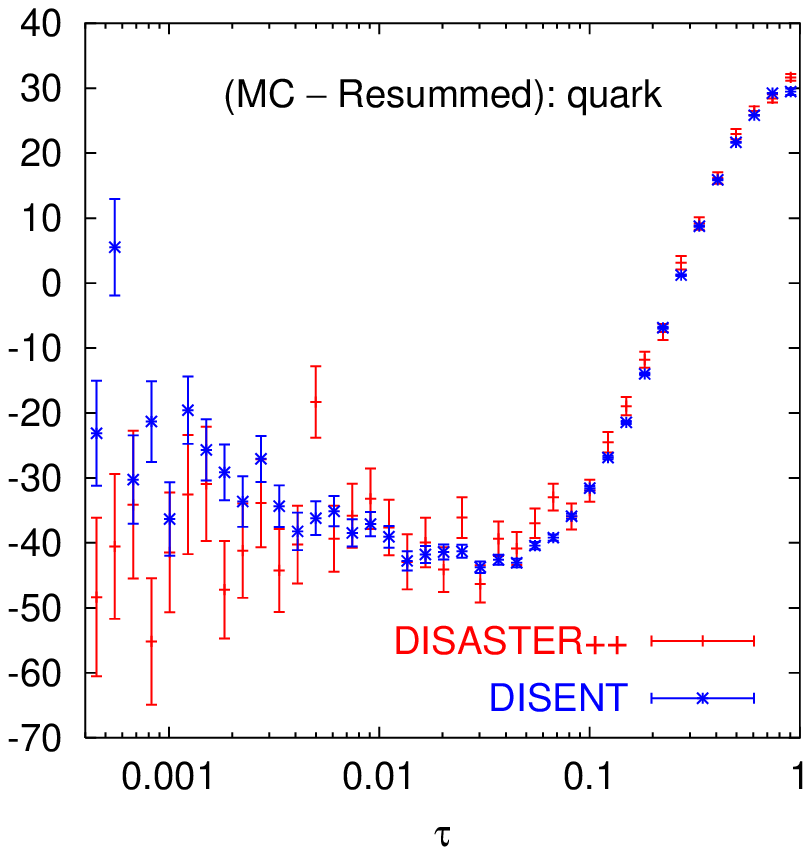,width=0.43\textwidth}\quad
  \epsfig{file=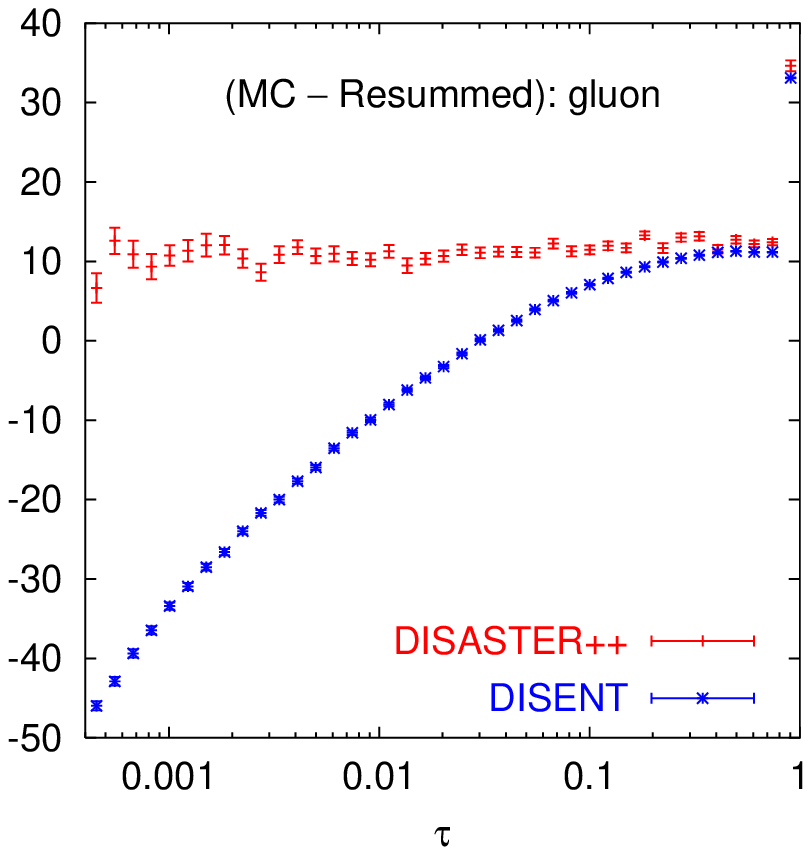,width=0.43\textwidth}
   \caption{Coefficient of $\left(\frac{\as}{2\pi}\right)^2$ in 
     $\frac{\tau}{\sigma_0} \frac{d\sigma}{d\tau}$. See text for
     details.}
  \label{fig:disdis}
  \end{center}
\end{figure}

Expanding \eqref{eq:sigtau} to NLO it is possible to perform a
comparison to the fixed order Monte Carlo programs DISASTER++
\cite{DR} and DISENT \cite{DT}. We actually look at the coefficient
of $(\as/2\pi)^2$ in
\begin{equation}
  \label{eq:lookat}
  \frac{\tau}{\sigma_0} \frac{d\sigma(\tau)}{d\tau}\,,
\end{equation}
where we have normalised to the the Born cross section $\sigma_0$ for
simplicity. At second order eq.~\eqref{eq:lookat} contains terms
$\ln^n \tau$, $n \le 3$ and for the resummation to be correct to
next-to-leading order we should correctly control terms $1 \le n \le
3$.  So the difference between our expanded result and the exact
result should at most be a constant for small $\tau$. The upper two
plots of figure~\ref{fig:disdis} show our results for the quark and
gluon-initiated components of the answer compared to the predictions
from DISASTER++. The shape of the distributions is well reproduced for
small $\tau$.  The lower two plots show the difference between the
DISASTER++ results and ours: one sees that for small $\tau$ this
difference is indeed compatible with a constant, as required.
Comparisons have also been made with DISENT, which seems to disagree
with our result in the gluon sector at the level of a term
proportional to $\ln^2 \tau$, and perhaps also in the quark sector.

\section{Outlook}
\label{sec:future}

Although we have results for the resummation of the thrust
distribution in DIS, some work remains to be done for the practical
implementation of our results for comparison with experimental data.
In particular, prescriptions need to be defined for the matching of
our resummed result with the fixed order results, in order extend the
range of applicability of the calculation (without the matching the
results are applicable only to very small $\tau$).

Work is also in progress on the resummation of other DIS event-shape
variables. Among these, the thrust normalised to the energy in the
current hemisphere is close to completion. Other variables to be
studied include the jet mass, the $C$-parameter and the jet-broadening
(the resummation of the latter is relevant also for predicting the
form of the power correction \cite{DMS}).

Once this programme is complete, we hope that it will lead to
phenomenological studies analogous to those already being carried out
in $\ee$.

\section*{Acknowledgements}
We benefited much from continuous discussions of this and related
subjects with Yuri Dokshitzer, Pino Marchesini and Mike Seymour.  One
of us (VA) is grateful to Yuri Dokshitzer and Pino Marchesini for
introducing him to perturbative QCD, and to the organisers and the
participants of the 1999 U.K. Phenomenology Workshop on Collider
Physics for the invitation and for the stimulating atmosphere. He
would also like to thank N. Brook and V. Khoze for useful discussions.
This research was supported in part by MURST, Italy and by the EU
Fourth Framework Programme `Training and Mobility of Researchers',
Network `Quantum Chromodynamics and the Deep Structure of Elementary
Particles', contract FMRX-CT98-0194 (DG 12-MIHT).

\end{document}